\documentclass[copyright,creativecommons]{eptcs}
\providecommand{\event}{GANDALF 2010} 
\usepackage{breakurl}             
\usepackage{amssymb}
\usepackage{color}
\usepackage{graphicx}
\usepackage{program}
\usepackage{url}

\normalbaroutside

\title{Using Strategy Improvement to Stay Alive\thanks{This work has
been partially supported by the Grant Agency of the Czech Republic
grants No. 201/09/1389, 102/09/H042.}}
\author{Lubo\v{s} Brim
\institute{Masaryk University\\ Brno, Czech Republic}
\email{brim@fi.muni.cz}
\and
Jakub Chaloupka
\institute{Masaryk University\\ Brno, Czech Republic}
\email{xchalou1@fi.muni.cz}
}
\def\titlerunning{Using Strategy Improvement to Stay Alive}
\def\authorrunning{L. Brim, J. Chaloupka}
\begin{document}
\maketitle

\bibliographystyle{eptcs}

\newcommand{\G}{G}
\newcommand{\Max}{\textnormal{\scriptsize Max}}
\newcommand{\Min}{\textnormal{\scriptsize Min}}
\newcommand{\pt}{\ensuremath{\mathsf{P}}}
\newcommand{\np}{\ensuremath{\mathsf{NP}}}
\newcommand{\up}{\ensuremath{\mathsf{UP}}}
\newcommand{\co}[1]{\ensuremath{\mathsf{co}\mbox{--}#1}}
\newcommand{\lb}{\textsf{lb}}
\newcommand{\lwub}{\textsf{lwub}}
\newcommand{\outcome}{\textsf{outcome}}
\newcommand{\ppath}{\textsf{path}}
\newcommand{\ppref}{\textsf{prefix}}
\newcommand{\psuff}{\textsf{suffix}}
\newcommand{\pseg}{\textsf{segment}}
\newcommand{\cycle}{\textsf{cycle}}

\newtheorem{theorem}{Theorem}
\newtheorem{lemma}{Lemma}
\newtheorem{example}{Example}

\begin{abstract}
  We design a novel algorithm for solving Mean-Payoff Games (MPGs).
  Besides solving an MPG in the usual sense, our algorithm computes
  more information about the game, information that is important with
  respect to applications. The weights of the edges of an MPG can be
  thought of as a gained/consumed energy -- depending on the sign. For
  each vertex, our algorithm computes the minimum amount of initial
  energy that is sufficient for player Max to ensure that in a play
  starting from the vertex, the energy level never goes below zero.
  Our algorithm is not the first algorithm that computes the minimum
  sufficient initial energies, but according to our experimental study
  it is the fastest algorithm that computes them. The reason is that
  it utilizes the strategy improvement technique which is very
  efficient in practice.  
\end{abstract}

\section{Introduction}\label{sec:intro}

A \emph{Mean-Payoff Game (MPG)}
\cite{ehrenfeucht79positional,gurvich88cyclic,zwick96complexity} is a
two-player infinite game played on a finite weighted directed graph, the
vertices of which are divided between the two players. A play starts by
placing a token on some vertex and the players, named Max and Min, move the
token along the edges of the graph ad infinitum. If the token is on Max's
vertex, he chooses an outgoing edge and the token goes to the destination
vertex of that edge. If the token is on Min's vertex, it is her turn to
choose an outgoing edge. Roughly speaking, Max wants to maximize the average
weight of the traversed edges whereas Min wants to minimize it. It was
proved in \cite{ehrenfeucht79positional} that each vertex $v$ has a
\emph{value}, denoted by $\nu(v)$, which each player can secure by a
positional strategy, i.e., strategy that always chooses the same outgoing
edge in the same vertex. To solve an MPG is to find the values of all
vertices, and, optionally, also strategies that secure the values.

In this paper we deal with MPGs with other than the standard average-weight
goal. Player Max now wants the sum of the weights of the traversed edges,
plus some initial value (initial ``energy''), to be non-negative at each
moment of the play. He also wants to know the minimal sufficient amount of
initial energy that enables him to stay non-negative. For different starting
vertices, the minimal sufficient initial energy may be different and for
starting vertices with $\nu < 0$, it is impossible to stay non-negative with
arbitrarily large amount of initial energy.

The problem of computation of the minimal sufficient initial energies has
been studied under different names by Chakrabarti et
al.~\cite{chakrabarti03resource}, Lifshits and Pavlov~\cite{lifshits06fast},
and Bouyer et al.~\cite{bouyer08infinite}. In~\cite{chakrabarti03resource}
it was called the problem of \emph{pure energy interfaces},
in~\cite{lifshits06fast} it was called the problem of \emph{potential
  computation}, and in~\cite{bouyer08infinite} it was called \emph{the
  lower-bound problem}. The paper~\cite{bouyer08infinite} also contains the
definition of a similar problem -- \emph{the lower-weak-upper-bound
  problem}. An instance of this problem contains, besides an MPG, also a
bound $b$. The goal is the same, Max wants to know how much initial energy
he needs to stay non-negative forever, but now the energy level is bounded
from above by $b$ and during the play, all increases above this bound are
immediately truncated.

Various resource scheduling problems for which the standard solution of an
MPG is not useful can be formulated as the lower-bound or the
lower-weak-upper-bound problems, which extends the applicability of
MPGs. For example, an MPG can be used to model a robot in a hostile
environment. The weights of edges represent changes in the remaining battery
capacity of the robot -- positive edges represent recharging, negative edges
represent energy consuming actions. The bound $b$ is the maximum capacity of
the battery. Player Max chooses the actions of the robot and player Min
chooses the actions of the hostile environment. By solving the
lower-weak-upper-bound problem, we find out if there is some strategy of the
robot that allows him to survive in the hostile environment, i.e., its
remaining battery capacity never goes below zero, and if there is such a
strategy, we also get the minimum initial remaining battery capacity that
allows him to survive.

The first algorithm solving the lower-bound problem was proposed by
Chakrabarti et al.~\cite{chakrabarti03resource} and it is based on value
iteration. 
The algorithm can also be easily modified to solve the
lower-weak-upper-bound problem. The value iteration algorithm was later
improved by Chaloupka and Brim in~\cite{chaloupka09faster}, and
independently by Doyen, Gentilini, and Raskin~\cite{doyen09faster}, extended
version of~\cite{chaloupka09faster, doyen09faster} was recently
submitted~\cite{brim10faster}. Henceforward we will use the term ``value
iteration'' (VI) to denote only the improved version
from~\cite{chaloupka09faster, doyen09faster}. The algorithms of Bouyer et
al.~\cite{bouyer08infinite} that solve the two problems are essentially the
same as the original algorithm from~\cite{chakrabarti03resource}. However,
\cite{bouyer08infinite} focuses mainly on other problems than the
lower-bound and the lower-weak-upper-bound problems for MPGs.  Different
approach to solving the lower-bound problem was proposed by
Lifshits~and~Pavlov~\cite{lifshits06fast}, but their algorithm has
exponential space complexity, and so it is not appropriate for practical
use. VI seems to be the best known approach to solving the two problems.

In this paper, we design a novel algorithm based on the strategy improvement
technique, suitable for practical solving of the lower-bound and the
lower-weak-upper-bound problems for large MPGs. The use of the strategy
improvement technique for solving MPGs goes back to the algorithm of Hoffman
and Karp from~1966~\cite{hoffman66nonterminating}. Their algorithm can be
used to solve only a restricted class of MPGs, but strategy improvement
algorithms for solving MPGs in general exist as
well~\cite{bjorklund07combinatorial,schewe08optimal,cochetterrasson06policy}.
However, all of them solve neither the lower-bound nor the
lower-weak-upper-bound problem (cf. Section~\ref{sec:xp}, first part, last
paragraph), our algorithm is the first. Another contribution of this paper
is a further improvement of VI.

The shortcoming of VI is that it takes enormous time on MPGs with at least
one vertex with $\nu < 0$. Natural way to alleviate this problem is to find
the vertices with $\nu < 0$ by some fast algorithm and run VI on the
rest. Based on our previous experience with algorithms for solving
MPGs~\cite{chaloupka09parallel}, we selected two algorithms for computation
of the set of vertices with $\nu < 0$. Namely, the algorithm of
Bj\"{o}rklund and Vorobyov~\cite{bjorklund07combinatorial} (BV), and the
algorithm of Schewe~\cite{schewe08optimal} (SW). This gives us two
algorithms: VI~+~BV and VI~+~SW. However, the preprocessing is not helpful
on MPGs with all vertices with $\nu \geq 0$, and it is also not helpful for
solving the lower-weak-upper-bound problem for small bound $b$. Therefore,
we also study the algorithm VI without the preprocessing.

Our new algorithm based on the strategy improvement technique that we
propose in this paper has the complexity $O(|V| \cdot (|V| \cdot \log |V| +
|E|) \cdot W)$, where $W$ is the maximal absolute edge-weight. It is
slightly worse than the complexity of VI, the same as the complexity of
VI~+~BV, and better than the complexity of VI~+~SW. We call our algorithm
``Keep Alive Strategy Improvement'' (KASI). It solves both the lower-bound
and the lower-weak-upper-bound problem. Moreover, as each algorithm that
solves the lower-bound problem also divides the vertices of an MPG into
those with $\nu \geq 0$ and those with $\nu < 0$, which can be used to
compute the exact $\nu$ values of all vertices, KASI can be thought of as an
algorithm that also solves MPGs in the usual sense.  As a by-product of the
design of KASI, we improved the complexity of BV and proved that Min may not
have positional strategy that is also optimal with respect to the
lower-weak-upper-bound problem. Moreover, we describe a way to construct an
optimal strategy for Min with respect to the lower-weak-upper-bound problem.

To evaluate and compare the algorithms VI, VI~+~BV, VI~+~SW, and KASI, we
implemented them and carried out an experimental study. According to the
study, KASI is the best algorithm.

\section{Preliminaries}\label{sec:pre}

A \emph{Mean-Payoff Game} (MPG)
\cite{ehrenfeucht79positional,gurvich88cyclic,zwick96complexity} is given by
a triple $\Gamma = (\G, V_\Max, V_\Min)$, where $\G = (V, E, w)$ is a finite weighted
directed graph such that $V$ is a disjoint union of the sets $V_\Max$ and
$V_\Min$, $w : E \rightarrow \mathbb{Z}$ is the weight function, and each $v
\in V$ has out-degree at least one.  The game is played by two opposing
players, named Max and Min. A play starts by placing a token on some given
vertex and the players then move the token along the edges of $\G$ ad
infinitum. If the token is on vertex $v \in V_\Max$, Max moves it. If the
token is on vertex $v \in V_\Min$, Min moves it. This way an infinite path
$p = (v_0, v_1, v_2, \dots)$ is formed. Max's aim is to maximize his gain:
$\liminf_{n \rightarrow \infty} \frac{1}{n} \sum_{i=0}^{n-1} w(v_i,
v_{i+1})$, and Min's aim is to minimize her loss: $\limsup_{n \rightarrow
  \infty} \frac{1}{n} \sum_{i=0}^{n-1} w(v_i, v_{i+1})$. For each vertex $v
\in V$, we define its \emph{value}, denoted by $\nu(v)$, as the maximal gain
that Max can ensure if the play starts at vertex $v$. It was proved that it
is equal to the minimal loss that Min can ensure. Moreover, both players can
ensure $\nu(v)$ by using positional strategies defined below
\cite{ehrenfeucht79positional}.

A (general) \emph{strategy} of Max is a function $\sigma : V^* \cdot V_\Max
\rightarrow V$ such that for each finite path $p = (v_0, \dots, v_k)$ with
$v_k \in V_\Max$, it holds that $(v_k, \sigma(p)) \in E$. Recall that each
vertex has out-degree at least one, and so the definition of a strategy is
correct. The set of all strategies of Max in $\Gamma$ is denoted by
$\Sigma^\Gamma$. We say that an infinite path $p = (v_0, v_1, v_2, \dots)$
agrees with the strategy $\sigma \in \Sigma^\Gamma$ if for each $v_i \in
V_\Max$, $\sigma(v_0, \dots, v_i) = v_{i + 1}$. A strategy $\pi$ of Min is
defined analogously. The set of all strategies of Min in $\Gamma$ is denoted
by $\Pi^\Gamma$. Given an initial vertex $v \in V$, the \emph{outcome} of
two strategies $\sigma \in \Sigma^\Gamma$ and $\pi \in \Pi^\Gamma$ is the
(unique) infinite path $\outcome^{\Gamma}(v, \sigma, \pi) = (v = v_0, v_1,
v_2, \dots)$ that agrees with both $\sigma$ and $\pi$.

The strategy $\sigma \in \Sigma^\Gamma$ is called a \emph{positional
  strategy} if $\sigma(p) = \sigma(p')$ for all finite paths $p = (v_0,
\dots, v_k)$ and $p' = (v'_0, \dots, v'_{k'})$ such that $v_k = v'_{k'} \in
V_\Max$. For the sake of simplicity, we think of a positional strategy of
Max as a function $\sigma : V_\Max \rightarrow V$ such that $(v, \sigma(v))
\in E$, for each $v \in V_\Max$. The set of all positional strategies of Max
in $\Gamma$ is denoted by $\Sigma_M^\Gamma$. A positional strategy $\pi$ of
Min is defined analogously. The set of all positional strategies of Min in
$\Gamma$ is denoted by $\Pi_M^\Gamma$. We define $\G_\sigma$, the
\emph{restriction of $\G$ to $\sigma$}, as the graph $(V, E_\sigma,
w_\sigma)$, where $E_\sigma = \{ (u, v) \in E \mid u \in V_\Min \vee
\sigma(u) = v\}$, and $w_\sigma = w \restriction {E_\sigma}$. That is, we get
$\G_\sigma$ from $\G$ by deleting all the edges emanating from Max's
vertices that do not follow $\sigma$. $\G_\pi$ for a strategy $\pi$ of Min
is defined analogously. For $\sigma \in \Sigma_M^\Gamma$, we also define
$\Gamma_\sigma = (\G_\sigma, V_\Max, V_\Min)$, and for $\pi \in
\Pi_M^\Gamma$, $\Gamma_\pi = (\G_\pi, V_\Max, V_\Min)$.

The \emph{lower-bound problem} for an MPG $\Gamma = (\G = (V, E, w), V_\Max,
V_\Min)$ is the problem of finding $\lb^\Gamma(v) \in \mathbb{N}_0 \cup
\{\infty\}$ for each $v \in V$, such that: 

$$
\begin{array}[t]{lll}
\lb^\Gamma(v) = \min\{ x \in \mathbb{N}_0 \mid {} & 
\multicolumn{2}{l}{(\exists \sigma \in \Sigma^\Gamma)(\forall \pi \in \Pi^\Gamma)} \\
& ( & \outcome^\Gamma(v, \sigma, \pi) = (v = v_0, v_1, v_2,\dots) \wedge {} \\
& & (\forall n \in \mathbb{N})(x + \sum_{i=0}^{n-1} w(v_i, v_{i+1}) \geq 0)
\ )\ \}\end{array}
$$

where minimum of an empty set is $\infty$. That is, $\lb^\Gamma(v)$ is the
minimal sufficient amount of initial energy that enables Max to keep the
energy level non-negative forever, if the play starts from $v$. If
$\lb^\Gamma(v) = \infty$, which means that $\nu(v) < 0$, then we say that
Max loses from~$v$, because arbitrarily large amount of initial energy is
not sufficient. If $\lb^\Gamma(v) \in \mathbb{N}_0$, then Max wins from $v$.

The strategy $\sigma \in \Sigma^\Gamma$ is an \emph{optimal strategy of Max with
  respect to the lower-bound problem}, if it ensures that for each $v \in V$
such that $\lb^\Gamma(v) \neq \infty$, $\lb^\Gamma(v)$ is a sufficient
amount of initial energy. 

The strategy $\pi \in \Pi^\Gamma$ is an \emph{optimal strategy of Min with
  respect to the lower-bound problem}, if it ensures that for each $v \in V$
such that $\lb^\Gamma(v) \neq \infty$, Max needs at least $\lb^\Gamma(v)$
units of initial energy, and for each $v \in V$ such that $\lb^\Gamma(v) =
\infty$, Max loses.

The \emph{lower-weak-upper-bound problem} for an MPG $\Gamma = (\G = (V, E,
w), V_\Max, V_\Min)$ and a bound $b \in \mathbb{N}_0$ is the problem of
finding $\lwub_b^\Gamma(v) \in \mathbb{N}_0 \cup \{\infty\}$ for each $v \in
V$, such that:

$$
\begin{array}[t]{lll}
  \lwub_b^\Gamma(v) = \min\{ x \in \mathbb{N}_0 \mid {} & 
  \multicolumn{2}{l}{(\exists \sigma \in \Sigma^\Gamma)(\forall \pi \in \Pi^\Gamma)} \\
  & ( & \outcome^\Gamma(v, \sigma, \pi) = (v = v_0, v_1, v_2,\dots) \wedge {} \\
  & & (\forall n \in \mathbb{N})(x + \sum_{i=0}^{n-1} w(v_i, v_{i+1}) \geq 0) \wedge {}  \\
  & & (\forall n_1, n_2 \in \mathbb{N}_0)(n_1 < n_2 \Rightarrow 
  \sum_{i=n_1}^{n_2-1} w(v_i, v_{i+1}) \geq -b)\ )\ \} \end{array}
$$

where minimum of an empty set is $\infty$. That is, $\lwub_b^\Gamma(v)$ is
the minimal sufficient amount of initial energy that enables Max to keep the
energy level non-negative forever, if the play starts from $v$, under the
additional condition that the energy level is truncated to $b$ whenever it
exceeds the bound. The additional condition is equivalent to the condition
that the play does not contain a segment of weight less than $-b$. If
$\lwub_b^\Gamma(v) = \infty$, then we say that Max loses from $v$, because
arbitrarily large amount of initial energy is not sufficient.  If
$\lwub_b^\Gamma(v) \in \mathbb{N}_0$, then Max wins from $v$. Optimal
strategies for Max and Min with respect to the lower-weak-upper-bound
problem are defined in the same way as for the lower-bound problem.

It was proved in~\cite{bouyer08infinite} that both for the lower-bound
problem and the lower-weak-upper-bound problem Max can restrict himself only
to positional strategies, i.e., he always has a positional strategy that is
also optimal. Therefore, we could use the set $\Sigma_M^\Gamma$ instead of
the set $\Sigma^\Gamma$ in the definitions of both the lower-bound problem
and the lower-weak-upper-bound problem.

In the rest of the paper, we will focus only on the lower-weak-upper-bound
problem, because it includes the lower-bound problem as a special case. The
reason is that for each $v \in V$ such that $\lb^\Gamma(v) < \infty$, it
holds that $\lb^\Gamma(v) \leq (|V| - 1)\cdot W$, where $W$ is the maximal
absolute edge-weight in $\G$. It was proved
in~\cite{bouyer08infinite}. Therefore, if we choose $b = (|V| - 1)\cdot W$,
then for each $v \in V$, $\lb^\Gamma(v) = \lwub_b^\Gamma(v)$.

Let $\G = (V, E, w)$ be a weighted directed graph, let $p = (v_0, \dots,
v_k)$ be a path in $\G$, and let $c = (u_0, \dots, u_{r - 1}, u_r = u_0)$ be a
cycle in $\G$. Then $w(p)$, the weight of $p$, and $w(c)$, the weight of $c$,
are defined in the following way: $w(p) = \sum_{i = 0}^{k - 1} w(v_i, v_{i +
  1})$, $w(c) = \sum_{i = 0}^{r - 1} w(u_i, u_{i + 1})$.

Let $\Gamma = (\G = (V, E, w), V_\Min, V_\Max)$ be an MPG and let $D
\subseteq V$. Then $\G(D)$ is the subgraph of $\G$ induced by the set
$D$. Formally, $\G(D) = (D, E \cap D \times D, w \restriction D \times
D)$. We also define the restriction of $\Gamma$ induced by $D$. Since some
vertices might have zero out-degree in $\G(D)$, we define $\Gamma(D) =
(\G'(D), V_\Min \cap D, V_\Max \cap D)$, where $\G'(D)$ is the graph $\G(D)$
with negative self-loops added to all vertices with zero outdegree. That is,
we make the vertices with zero out-degree in $\G(D)$ losing for Max in
$\Gamma(D)$ with respect to the the lower-weak-upper-bound problem.

Let $\G = (V, E, w)$ be a graph and let $B, A \subseteq V$. If we say that
``$p$ is a path from $v$ to $B$'' we mean a path with the last vertex and
only the last vertex in $B$, formally: $p = (v = v_0, \dots, v_k)$, where
$v_0, \dots, v_{k - 1} \in V \setminus B \wedge v_k \in B$. Furthermore, a
path from $A$ to $B$ is a path from $v$ to $B$ such that $v \in A$. The term
``longest'' in connection with paths always refers to the weights of the
paths, not the numbers of edges.

Operations on vectors of the same dimension are element-wise. For example,
if $d_0$ and $d_1$ are two vectors of dimension $|V|$, then $d_0 < d_1$
means that for each $v \in V$, $d_0(v) \leq d_1(v)$, and for some $v \in V$,
$d_0(v) < d_1(v)$.

For the whole paper let $\Gamma = (\G = (V, E, w), V_\Max, V_\Min)$ be an
MPG and let $W$ be the maximal absolute edge-weight in $\G$, i.e., $W = \max_{e
  \in E}|w(e)|$.

\section{The Algorithm}\label{sec:alg}

High-level description of our Keep Alive Strategy Improvement algorithm
(KASI) for the lower-weak-upper-bound problem is as follows.  Let $(\Gamma,
b)$ be an instance of the lower-weak-upper-bound problem. KASI maintains a
vector $d \in (\mathbb{Z} \cup \{-\infty\})^V$ such that $-d \geq {\bf 0}$
is always a lower estimate of $\lwub^\Gamma_b$, i.e., $-d \leq
\lwub^\Gamma_b$. The vector $d$ is gradually decreased, and so $-d$ is
increased, until $-d = \lwub^\Gamma_b$. The reason why KASI maintans the
vector $d$ rather than $-d$ is that $d$ contains weights of certain paths
and we find it more natural to keep them as they are, than to keep their
opposite values. The algorithm also maintains a set $D$ of vertices such
that about the vertices in $V \setminus D$ it already knows that they have
infinite $\lwub_b^\Gamma$ value. Initially, $d = {\bf 0}$ and $D = V$. KASI
starts with an arbitrary strategy $\pi \in \Pi_M^\Gamma$ and then
iteratively improves it until no improvement is possible. In each iteration,
the current strategy is first evaluated and then improved. The strategy
evaluation examines the graph $G_\pi(D)$ and updates the vector $d$ so that
for each $v \in D$, it holds \vspace{-1em}

$$-d(v) = \lwub_b^{\Gamma_\pi(D)}(v)$$

That is, it solves the lower-weak-upper-bound problem in the
restricted game $\Gamma_\pi(D)$, where Min has no choices. This
explains why the restricted game was defined the way it was, because
if a vertex from $D$ has outgoing edges only to $V \setminus D$, then
it is losing for Max in $\Gamma$. The vertices with the $d$~value
equal to $-\infty$ are removed from the set $D$. Since the strategy
$\pi$ is either the first strategy or an improvement of the previous
strategy, the vector $d$ is always decreased by the strategy
evaluation and we get a better estimate of $\lwub^\Gamma_b$. To
improve the current strategy the algorithm checks whether for some
$(v, u) \in E$ such that $v \in V_\Min$ and $d(v) > -\infty$ it holds
that $d(v) > d(u) + w(v, u)$.  This is called a \emph{strategy
  improvement condition}.  Such an edge indicates that $-d(v)$ is not
a sufficient initial energy at $v$, because traversing the edge $w(v,
u)$ and continuing from $u$ costs at least $- w(v, u) - d(u)$ units of
energy, which is greater than $-d(v)$ (Recall that $-d$ is a lower
estimate of $\lwub^\Gamma_b$). If there are edges satisfying the
condition, the strategy $\pi$ is improved in the following way. For
each vertex $v \in V_\Min$ such that there is an edge $(v, u) \in E$
such that $d(v) > d(u) + w(v, u)$, $\pi(v)$ is switched to $u$. If $v$
has more than one such edge emanating from it, any of them is
acceptable. Then, another iteration of KASI is started. If no such
edge exists, the algorithm terminates, because it holds that each
vertex $v \in V$ has $-d(v) = \lwub_b^\Gamma(v)$. Detailed description
of the algorithm follows.

In Figure~\ref{fig:eval} is a pseudo-code of the strategy evaluation part of
our algorithm. The input to the procedure consists of four parts. The first
and the second part form the lower-weak-upper-bound problem instance that
the main algorithm KASI is solving, the MPG $\Gamma$ and the bound $b \in
\mathbb{N}_0$. The third part is the strategy $\pi \in \Pi_M^\Gamma$ that we
want to evaluate and the fourth part of the input is a vector $d_{-1} \in
(\mathbb{Z} \cup \{-\infty\})^{V}$.  The vector $d_{-1}$ is such that
$-d_{-1}$ is a lower estimate of $\lwub^\Gamma_b$, computed for the previous
strategy, or, in case of the first iteration of KASI, set by initialization
to a vector of zeros. Let $A = \{v \in V \mid d_{-1}(v) = 0\}$ and $D = \{v
\in V \mid d_{-1}(v) > -\infty\}$, then the following conditions hold.

\begin{enumerate}
\item[i.] Each cycle in $\G_\pi(D\setminus A)$ is negative
\item[ii.] For each $v \in D \setminus A$, it holds that $d_{-1}(v) < 0$
      and for each edge $(v, u) \in E_\pi$, i.e., for each edge
      emanating from $v$ in $\G_\pi$, it holds that $d_{-1}(v) \geq
      d_{-1}(u) + w(v, u)$.
\end{enumerate}

From these technical conditions it follows that $-d_{-1}$ is also a lower
estimate of $\lwub^{\Gamma_\pi(D)}_b$ and the purpose of the strategy
evaluation procedure is to decrease the vector $d_{-1}$ so that the
resulting vector~$d$ satisfies $-d = \lwub^{\Gamma_\pi(D)}_b$. To see why
from (i.) and (ii.) it follows that $-d_{-1} \leq \lwub^{\Gamma_\pi(D)}_b$,
consider a path $p = (v_0, \dots, v_k)$ from $D \setminus A$ to $A$ in
$G_\pi(D)$. From (ii.) it follows that for each $j \in \{0, \dots, k - 1\}$,
it holds that $d_{-1}(v_j) \geq d_{-1}(v_{j + 1}) + w(v_j, v_{j + 1})$. If
we sum the inequalities, we get $d_{-1}(v_0) \geq d_{-1}(v_k) + w(p)$. Since
$v_k \in A$, $d_{-1}(v_k) = 0$ and the inequality becomes $d_{-1}(v_0) \geq
w(p)$.  Therefore, each infinite path in $G_\pi(D)$ starting from $v \in D$
and containing a vertex from $A$ has a prefix of weight less or equal to
$d_{-1}(v_0)$. Furthermore, if the infinite path does not contain a vertex
from $A$, weights of its prefixes cannot even be bounded from below, because
by (i.), all cycles in $G_\pi(D \setminus A)$ are negative. All in all,
$-d_{-1}$ is a lower estimate of $\lwub^{\Gamma_\pi(D)}_b$.

The conditions (i.) and (ii.) trivially hold in the first iteration of
the main algorithm, for $d_{-1} = {\bf 0}$. In each subsequent
iteration, $d_{-1}$ is taken from the output of the previous iteration
and an intuition why the conditions hold will be given below.

The output of the strategy evaluation procedure is a vector $d \in
(\mathbb{Z} \cup \{-\infty\})^{V}$ such that for each $v \in D$, it holds
that $-d(v) = \lwub_b^{\Gamma_\pi(D)}(v)$. Recall that $D = \{v \in V \mid
d_{-1}(v) > -\infty\}$.

The strategy evaluation works only with the restricted graph
$\G_\pi(D)$ and it is based on the fact that if we have the set $B_z =
\{v \in D \mid \lwub_b^{\Gamma_\pi(D)}(v) = 0\}$, i.e., the set of
vertices where Max does not need any initial energy to win, then we
can compute $\lwub_b^{\Gamma_\pi(D)}$ of the remaining vertices by
computing longest paths to the set $B_z$. More precisely, for each
vertex $v \in D \setminus B_z$, $\lwub_b^{\Gamma_\pi(D)}(v)$ is equal to
the absolute value of the weight of a longest path from $v$ to $B_z$ in
$\G_\pi(D)$ such that the weight of each suffix of the path is greater
or equal to $-b$. If each path from $v$ to $B_z$ has a suffix of weight
less than $-b$ or no path from $v$ to $B_z$ exists, then
$\lwub_b^{\Gamma_\pi(D)}(v) = \infty$.

To get some idea about why this holds consider a play winning for Max.  The
energy level never drops below zero in the play, and so there must be a
moment from which onwards the energy level never drops below the energy
level of that moment. Therefore, Max does not need any initial energy to
win a play starting from the appropriate vertex (Please note that Min has no
choices in $\Gamma_\pi(D)$), and so $B_z$ is not empty. For the vertices in
$D \setminus B_z$, in order to win, Max has to get to some vertex in $B_z$
without exhausting all of his energy. So the minimal sufficient energy to
win is the minimal energy that Max needs to get to some vertex in $B_z$.
All paths from $D \setminus B_z$ to $B_z$ must be negative (otherwise $B_z$
would be larger), and so the minimal energy to get to $B_z$ is the absolute
value of the weight of a longest path to $B_z$ such that the weight of each
suffix of the path is greater or equal to $-b$. If no path to $B_z$ exists
or all such paths have suffixes of weight less than $-b$, Max cannot win.

Initially, the procedure over-approximates the set $B_z$ by the set
$B_0$ of vertices $v$ with $d_{-1}(v) = 0$ that have an edge $(v, u)$
such that $w(v, u) - d_{-1}(v) + d_{-1}(u) \geq 0$ emanating from them
(line~2), and then iteratively removes vertices from the set until it
arrives at the correct set $B_z$. The vector $-d_i$ is always a lower
estimate of $\lwub_b^{\Gamma_\pi(D)}$, i.e., it always holds that
$-d_i \leq \lwub_b^{\Gamma_\pi(D)}$. Therefore, only vertices $v$ with
$d_i(v) = 0$ are candidates for the final set $B_z$. However, the
vertices $v$ with $d_i(v) = 0$ such that for each edge $(v, u)$, it
holds that $w(v, u) - d_i(v) + d_i(u) < 0$ are removed from the set of
candidates. The reason is that since $d_i(v) = 0$, the inequality can
be developed to $-w(v, u) - d_i(u) > 0$, and so if the edge $(v, u)$
is chosen in the first step, then more than zero units of initial
energy are needed at $v$. During the execution of the procedure, $d_i$
decreases, and so $-d_i$ increases, until $-d_i =
\lwub_b^{\Gamma_\pi(D)}$.

In each iteration, the procedure uses a variant of the Dijkstra's algorithm
to compute longest paths from all vertices to $B_i$ on line~4. Since $B_i$
is an over-approximation of $B_z$, the absolute values of the weights of the
longest paths are a lower estimate of $\lwub_b^{\Gamma_\pi(D)}$. The weights
of the longest paths are assigned to $d_i$. In particular, for each $v \in
B_i$, $d_i(v) = 0$.  Dijkstra's algorithm requires all edge-weights be
non-positive (Please note that we are computing longest paths). Since
edge-weights are arbitrary integers, we apply potential transformation on
them to make them non-positive.  As vertex potentials we use $d_{i - 1}$,
which contains the longest path weights computed in the previous iteration,
or, in case $i = 0$, is given as input.  Transformed weight of an edge $(x,
y)$ is $w(x, y) - d_{i - 1}(x) + d_{i - 1}(y)$, which is always non-positive
for the relevant edges. In the first iteration of the main algorithm it
follows from the condition (ii.), and in the subsequent iterations it
follows from properties of longest path weights and the fact that only
vertices with all outgoing edges negative with the potential transformation
are removed from the candidate set.

The Dijkstra's algorithm is also modified so that it assigns $-\infty$ to
each $v \in D$ such that each path from $v$ to $B_i$ has a suffix of weight
less than $-b$.  Therefore, the vertices from which $B_i$ is not reachable
or is reachable only via paths with suffixes of weight less than $-b$ have
$d_i$ equal to $-\infty$. Also, vertices from $V \setminus D$ have $d_i$
equal to $-\infty$. A detailed description of $\textsc{Dijkstra}()$ is in
the full version of this paper~\cite{brim10usingTR}.

On line~5, the variable $i$ is increased (thus the current longest path
weights are now in $d_{i - 1}$), and on line~6, we remove from $B_{i-1}$
each vertex $v$ that does not have an outgoing edge $(v, u)$ such that $w(v,
u) - d_{i - 1}(v) + d_{i - 1}(u) \geq 0$. Another iteration is started only
if $B_i \neq B_{i - 1}$. If no vertex is removed on line~6, then $B_i = B_{i
  - 1}$ and the algorithm finishes and returns $d_{i - 1}$ as output. The
following theorem establishes the correctness of the algorithm. An intuition
why the theorem holds was given above. Its formal proof is in the full
version of this paper~\cite{brim10usingTR}.

\NumberProgramstrue
\begin{figure}[t]
\begin{program}
\PROC \textsc{EvaluateStrategy}(\Gamma, b, \pi, d_{-1})
  i := 0; B_0 := \{v \in V \mid d_{-1}(v) = 0 \wedge \max_{(v, u) \in E_\pi} (w(v, u) - d_{-1}(v) + d_{-1}(u)) \geq 0\}
  \WHILE i = 0 \OR B_{i-1} \neq B_i \DO
    d_i := \textsc{Dijkstra}(\G_\pi, b, B_i, d_{i - 1})
    i := i + 1
    B_i := B_{i-1} \setminus \{v \mid \max_{(v, u) \in E_\pi} (w(v, u)- d_{i - 1}(v) + d_{i - 1}(u)) < 0\}
  \OD
  \keyword{return}\ d_{i - 1}
\END
\end{program}
\vspace{-0.5cm}
\caption{Evaluation of strategy}\label{fig:eval}
\end{figure}

\begin{theorem}\label{thmb:computex} 
  Let $(\Gamma, b)$ be an instance of the lower-weak-upper-bound problem.
  Let further $\pi \in \Pi_M^\Gamma$ be a positional strategy of Min, and
  finally let $d_{-1} \in (\mathbb{Z} \cup \{-\infty\})^V$ be such that for
  $A = \{v \in V \mid d_{-1}(v) = 0\}$ and $D = \{v \in V \mid d_{-1}(v) >
  -\infty\}$, the conditions (i.) and (ii.) hold.
  Then for $d := \textsc{EvaluateStrategy}(\Gamma, b, \pi, d_{-1})$ it holds
  that for each $v \in D$, $d(v) = -\lwub_b^{\Gamma_\pi(D)}(v)$.
\end{theorem}

The complexity of $\textsc{EvaluateStrategy()}$ is $O(|V| \cdot (|V|\cdot
\log |V| + |E|))$. Each iteration takes $O(|V|~\cdot~\log |V|~+~|E|)$
because of $\textsc{Dijkstra()}$ and the number of iterations of the while
loop on lines~3--7 is at most $|V|$, because $B_i \subseteq V$ loses at
least one element in each iteration.

\begin{figure}[t]
\begin{program}
  \PROC \textsc{LowerWeakUpperBound}(\Gamma, b)
    i := 0; \pi_0 := |Arbitrary strategy from | \Pi_M^\Gamma\label{lwub:initstr}
    d_{-1} := {\bf 0}\label{lwub:initd} 
    improvement := true
    \WHILE improvement \DO\label{lwub:whilestart}
      d_i := \textsc{EvaluateStrategy}(\Gamma, b, \pi_i, d_{i - 1})\label{lwub:eval}
      improvement := false\label{lwub:aftereval}
      i := i + 1\label{lwub:inci}
      \pi_i := \pi_{i - 1}
      \FOREACH v \in V_\Min \DO\label{lwub:impstrstart}
         \IF d_{i - 1}(v) > -\infty \THEN
            \FOREACH (v, u) \in E \DO
               \IF d_{i - 1}(v) > d_{i - 1}(u) + w(v, u) \THEN
                 \pi_i(v) := u; improvement := true                 
               \FI
            \OD
         \FI
      \OD\label{lwub:impstrend}
    \OD\label{lwub:whileend}
    \keyword{return}\ -d_{i - 1}\label{lwub:return}
  \END
\end{program}
\vspace{-0.5cm}
\caption{Solving the lower-weak-upper-bound problem}\label{fig:lwub}
\end{figure}

In Figure~\ref{fig:lwub} is a pseudo-code of our strategy improvement
algorithm for solving the lower-weak-upper-bound problem using
$\textsc{EvaluateStrategy()}$. The input to the algorithm is a
lower-weak-upper-bound problem instance $(\Gamma, b)$. The output of
the algorithm is the vector $\lwub_b^\Gamma$. The pseudo-code
corresponds to the high-level description of the algorithm given at
the beginning of this section.

The algorithm proceeds in iterations. It starts by taking an arbitrary
strategy from $\Pi_M^\Gamma$ on line~\ref{lwub:initstr}, and initializing
the lower estimate of $\lwub_b^\Gamma$ to vector of zeros on
line~\ref{lwub:initd}. Then it alternates strategy evaluation
(line~\ref{lwub:eval}) and strategy improvement
(lines~\ref{lwub:impstrstart}--\ref{lwub:impstrend}) until no improvement is
possible at which point the main while-loop on
lines~\ref{lwub:whilestart}--\ref{lwub:whileend} terminates and the final
$d$ vector is returned on line~\ref{lwub:return}. At that point, it holds
that for each $v \in V$, $-d_{i - 1}(v) = \lwub_b^\Gamma(v)$. The whole
algorithm KASI is illustrated on Example~\ref{ex:kasi}. The following lemmas
and theorem establish the correctness of the algorithm.

\newlength{\oldtabcolsep}
\setlength{\oldtabcolsep}{\tabcolsep}
\setlength{\tabcolsep}{0.5cm}
\begin{figure}[t]
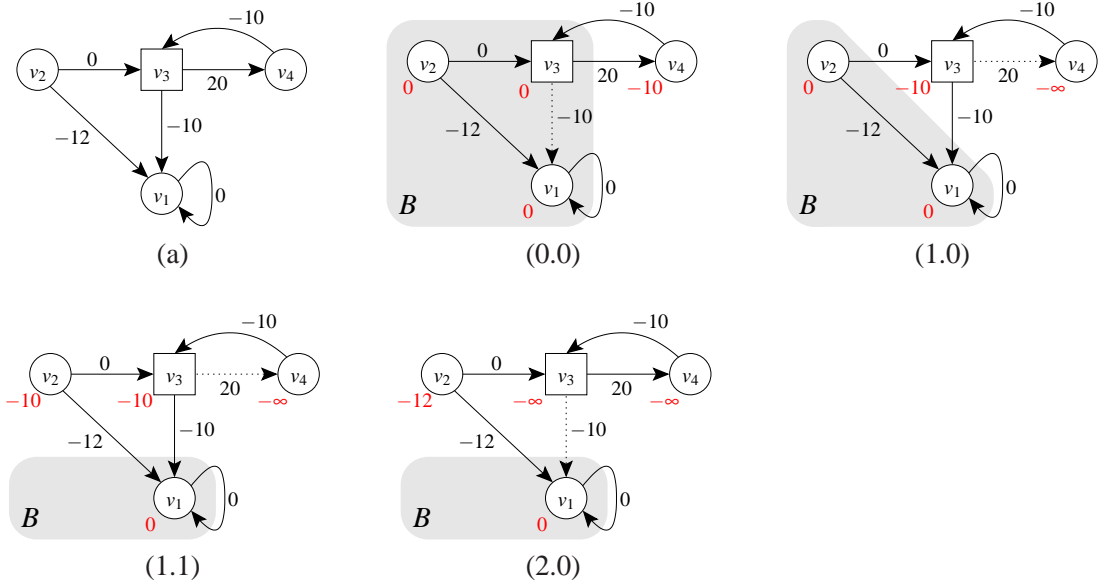

\begin{center}
\begin{tabular}{ccc}
  \hspace*{-0.3cm}\scalebox{0.55}{\input{kasi.pstex_t}} &
  \hspace*{-0.3cm}\scalebox{0.55}{\input{kasi1.pstex_t}} &
  \scalebox{0.55}{\input{kasi2.pstex_t}}\\  
  (a) & (0.0) & (1.0)\\
  & \\
  \hspace*{-0.3cm}\scalebox{0.55}{\input{kasi3.pstex_t}} & 
  \scalebox{0.55}{\input{kasi4.pstex_t}} & \\  
  (1.1) & (2.0) &\\
\end{tabular}
\end{center}
  \caption{Example of a Run of KASI}
  \label{fig:kasiex}
\end{figure}
\setlength{\tabcolsep}{\oldtabcolsep}

\begin{example}\label{ex:kasi}
  In Figure~\ref{fig:kasiex} is an example of a run of our algorithm KASI on
  a simple MPG. The MPG is in Figure~\ref{fig:kasiex}~(a). Circles are Max's
  vertices and the square is a Min's vertex. Let's denote the MPG by
  $\Gamma$, let $b = 15$ and consider a lower-weak-upper-bound problem given
  by $(\Gamma, b)$. Min has only two positional strategies, namely, $\pi^1$
  and $\pi^2$, where $\pi^1(v_3) = v_1$ and $\pi^2(v3) = v_4$. Let $\pi =
  \pi^2$ be the first selected strategy. For simplicity, we will use the
  symbols $\pi$, $d$, $B$, and $D$ without indices, although in pseudo-codes
  these symbols have indices, and the set $D$ of vertices with finite $d$
  value is not even explicitly used. Also, if we speak about a weight of an
  edge, we mean the weight with the potential transformation by
  $d$. Figure~\ref{fig:kasiex} illustrates the progress of the
  algorithm. Each figure denoted by $(r.s)$ shows a state of computation
  right after update of the vector $d$ by $\textsc{Dijkstra}()$. $r$ is the
  value of the iteration counter $i$ of $\textsc{LowerWeakUpperBound}()$,
  and $s$ is the value of the iteration counter $i$ of
  $\textsc{EvaluateStrategy}()$. In each figure, the $d$ value of each
  vertex is shown by that vertex. Edges that do not belong to the current
  strategy $\pi$ of Min are dotted. Detailed description of the progress of
  the algorithm follows. Initially, $\pi = \pi^2$, $d = {\bf 0}$, and $D =
  \{v_1, v_2, v_3, v_4\}$.

There are three vertices in $\G_\pi(D)$ with non-negative edges emanating
from them, namely, $v_1, v_2, v_3$, and so $\textsc{EvaluateStrategy}()$
takes $\{v_1, v_2, v_3\}$ as the first set $B$. After the vector $d$ is
updated so that it contains longest path weights to $B$
(Figure~\ref{fig:kasiex}~(0.0)), all vertices in $B$ still have non-negative
edges, and so the strategy evaluation finishes and the strategy improvement
phase is started. The strategy improvement condition is satisfied for the
edge $(v_3, v_1)$ and so $\pi$ is improved so that $\pi = \pi^1$. This
completes the first iteration of KASI and another one is started to evaluate
and possibly improve the new strategy $\pi$.

Now the vertex $v_3$ does not have a non-negative edge emanating from it, so
it is removed from the set $B$ and the longest path weights are recomputed
(Figure~\ref{fig:kasiex}~(1.0)). Please note that the only path from $v_4$ to
$B$ has a suffix of weight less than $-b$, and so $d(v_4) = -\infty$ and
$v_4$ is removed from the set $D$. The update to $d$ causes that $v_2$ does
not have a non-negative edge, thus it is also removed from the set $B$ and
the vector $d$ is recomputed again (Figure~\ref{fig:kasiex}~(1.1)). This
finishes the strategy evaluation and strategy improvement follows. The
strategy improvement condition is satisfied for the edge $(v_3, v_4)$, and
so the strategy $\pi^2$ is selected as the current strategy $\pi$
again. However, this is not the same situation as at the beginning, because
the set $D$ is now smaller. Evaluation of the strategy $\pi$ results in the
$d$ vector as depicted in Figure~\ref{fig:kasiex}~(2.0). The vertex $v_3$ has
$d(v_3) = -\infty$, because $v_3$ cannot reach the set $B$, which also
results in removal of $v_3$ from $D$. No further improvement of $\pi$ is
possible, and so $\lwub^\Gamma_b = -d = (0, 12, \infty, \infty)$.
\end{example}

\begin{lemma}\label{lemb:preterm}
  Every time line~\ref{lwub:eval} of $\textsc{LowerWeakUpperBound}()$ is
  reached, $\Gamma$, $b$, $\pi_i$, and $d_{i - 1}$ satisfy the assumptions
  of Theorem~\ref{thmb:computex}. Every time line~\ref{lwub:aftereval} of
  $\textsc{LowerWeakUpperBound}()$ is reached and $i > 0$, it holds that
  $d_i < d_{i - 1}$.
\end{lemma}

A formal proof of Lemma~\ref{lemb:preterm} is in the full version of this
paper~\cite{brim10usingTR}. The proof uses the following facts. The first
one we have already used: If $p$ is a path from $v$ to $u$ such that for
each edge $(x, y)$ in the path it holds that $d(x) \geq d(y) + w(x, y)$,
then $d(v) \geq d(u) + w(p)$, and if for some edge the inequality is strict,
then $d(v) > d(u) + w(p)$. The second fact is similar: If $c$ is a cycle
such that for each edge $(x, y)$ in the cycle it holds that $d(x) \geq d(y)
+ w(x, y)$, then $0 \geq w(c)$, and if for some edge the inequality is
strict, then the cycle is strictly negative. Using these facts we can now
give an intuition why the lemma holds.
 
The assumptions of Theorem~\ref{thmb:computex}, conditions (i.) and (ii.),
are trivially satisfied in the first iteration of
$\textsc{LowerWeakUpperBound}()$, as already mentioned. During execution
of $\textsc{EvaluateStrategy}()$, conditions (i.) and (ii.) remain
satisfied, for the following reasons. The $d$ values of vertices from $D$
are weights of longest paths to $B$, and so each edge $(x, y)$ emanating
from a vertex from $D \setminus B$ satisfies $d(x) \geq d(y) + w(x, y)$.
Only vertices with all outgoing edges negative with the potential
transformation are removed from the set $B$, i.e., only the vertices with
each outgoing edge $(x, y)$ satisfying $d(x) > d(y) + w(x, y)$.  Using the
facts from the previous paragraph, we can conclude that all newly formed
cycles in $\G_\pi(D \setminus B)$ are negative and the weights of longest
paths to $B$ cannot increase.  So to complete the intuition, it remains to
show why the conditions still hold after the strategy improvement and why
the strategy improvement results in decrease of the $d$ vector. This follows
from the fact that the new edges introduced by the strategy improvement are
negative with the potential transformation.

\begin{lemma}\label{lemb:zterm}
  The procedure $\textsc{LowerWeakUpperBound}()$ always terminates.
\end{lemma}

\proof By Lemma~\ref{lemb:preterm}, $d_i$ decreases in each iteration.  For
each $v \in V$, $d_i(v)$ is bounded from below by the term $-(|V| - 1) \cdot
W$, because it is the weight of some path in $\G$ with no repeated vertices
(Except for the case when $d_i(v) = -\infty$, but this is obviously not a
problem). Since $d_i$ is a vector of integers, infinite chain of
improvements is not possible, and so termination is guaranteed. \qed

Theorem~\ref{thmb:finito} is the main theorem of this paper which
establishes the correctness of our algorithm. Its proof is in the full
version of this paper~\cite{brim10usingTR}. The key idea of the proof is to
define strategies for both players with the following properties. Let $ds :=
\textsc{LowerWeakUpperBound}(\Gamma, b)$.  Max's strategy that we will
define ensures that for each vertex $v \in V$, $ds(v)$ is a sufficient
amount of initial energy no matter what his opponent does, and Min's
strategy that we will define ensures that Max cannot do with smaller amount
of initial energy. In particular, for vertices with $ds(v) = \infty$, the
strategy ensures that Max will eventually go negative or traverse a path
segment of weight less than $-b$ with arbitrarily large amount of initial
energy. From the existence of such strategies it follows that for each $v
\in V$, $ds(v) = \lwub_b^\Gamma(v)$, and both strategies are optimal with
respect to the lower-weak-upper-bound problem.

The optimal strategy of Max is constructed from the final longest path
forest computed by the procedure $\textsc{Dijkstra}()$ and the non-negative
(with potential transformation) edges emanating from the final set $B$. The
optimal strategy of Min is more complicated.

There is a theorem in~\cite{bouyer08infinite} which claims that Min can
restrict herself to positional strategies. Unfortunately, this is not
true. Unlike Max, Min sometimes needs memory. Example~\ref{ex:kasi} is a
proof of this fact, because none of the two positional strategies of Min
guarantees that Max loses from $v_3$. However, Min can play optimally using
the sequence of positional strategies computed by our algorithm. In
Example~\ref{ex:kasi}, to guarantee that Max loses from $v_3$, Min first
sends the play from $v_3$ to $v_4$ and when it returns back to $v_3$, she
sends the play to $v_1$. As a result, a path of weight $-20$ is traversed
and since $b = 15$, Max loses.

In general, let $\pi_0, \pi_1, \dots$ be the sequence of positional
strategies computed by the algorithm. Min uses the sequence in the following
way: if the play starts from a vertex with finite final $d$ value and never
leaves the set of vertices with finite final $d$ value, then Min uses the
last strategy in the sequence, and it is the best she can do, as stated by
Theorem~\ref{thmb:computex}. If the play starts or gets to a vertex with
infinite final $d$ value, she uses the strategy that caused that the $d$
value of that vertex became $-\infty$, but only until the play gets to a
vertex that was made infinite by some strategy with lower index. At that
moment Min switches to the appropriate strategy. In particular, Min never
switches to a strategy with higher index.

\begin{theorem}\label{thmb:finito} 
  Let $ds := \textsc{LowerWeakUpperBound}(\Gamma, b)$, then for each
  $v \in V$, $ds(v) = \lwub_b^\Gamma(v)$.
\end{theorem}

The algorithm has a pseudopolynomial time complexity: $O(|V|^2 \cdot (|V|
\cdot \log |V| + |E|) \cdot W)$. It takes $O(|V|^2 \cdot W)$ iterations
until the while-loop on lines~\ref{lwub:whilestart}--\ref{lwub:whileend}
terminates.  The reason is that for each $v \in V$, if $d(v) > -\infty$,
then $d(v) \geq -(|V| - 1) \cdot W$, because $d(v)$ is the weight of some
path with no repeated vertices, and so the $d$ vector can be improved at
most $O(|V|^2 \cdot W)$ times. Each iteration, if considered separately,
takes $O(|V| \cdot (|V|\cdot \log |V| + |E|))$, so one would say that the
overall complexity should be $O(|V|^3 \cdot (|V| \cdot \log |V| + |E|) \cdot
W)$.  However, the number of elements of the set $B_i$ in
$\textsc{EvaluateStrategy}()$ never increases, even between two distinct
calls of the evaluation procedure, hence the amortized complexity of one
iteration is only $O(|V|\cdot \log |V| + |E|)$. 

The algorithm can even be improved so that its complexity is $O(|V| \cdot
(|V| \cdot \log |V| + |E|) \cdot W)$. This is accomplished by efficient
computation of vertices which which will update their $d$ value in the next
iteration so that computational time is not wasted on vertices whose $d$
value is not going to change. Interestingly enough, the same technique can
be used to improve the complexity of the algorithm of Bj\"{o}rklund and
Vorobyov so that the complexities of the two algorithms are the
same. Detailed description of the technique is in the full version of this
paper~\cite{brim10usingTR}.

\section{Experimental Evaluation}\label{sec:xp}

Our experimental study compares four algorithms for solving the lower-bound
and the lower-weak-upper-bound problems. The first is value
iteration~\cite{chaloupka09faster, doyen09faster} (VI). The second and the
third are combinations of VI with other algorithm. Finally, the fourth
algorithm is our algorithm KASI. We will now briefly describe the algorithms
based on VI.

Let $(\Gamma, b)$ be an instance of the lower-weak-upper-bound problem. VI
starts with $d_0(v) = 0$, for each $v \in V$, and then computes $d_1, d_2,
\dots$ according to the following rules.

$$
d_{i + 1}(v) =
\left\{
\begin{array}{ll}
x = \min_{(v, u) \in E} \max (0, d_i(u) - w(v, u)) & \hspace{0.5cm}\textnormal{if } v \in V_\Max \wedge x \leq b\\ 
x = \max_{(v, u) \in E} \max (0, d_i(u) - w(v, u)) & \hspace{0.5cm}\textnormal{if } v \in V_\Min \wedge x \leq b\\
\infty & \hspace{0.5cm}\textnormal{otherwise}
\end{array} 
\right.
$$

It is easy to see that for each $v \in V$ and $k \in \mathbb{N}_0$,
$d_{k}(v)$ is the minimum amount of Max's initial energy that enables him to
keep the sum of traversed edges, plus $d_{k}(v)$, greater or equal to zero
in a $k$-step play. The computation continues until two consecutive $d$
vectors are equal. The last $d$ vector is then the desired vector
$\lwub_b^\Gamma$. If $b = (|V| - 1) \cdot W$, the algorithm solves the
lower-bound problem. The complexity of the straightforward implementation of
the algorithm is $O(|V|^2 \cdot |E| \cdot W)$, which was improved
in~\cite{chaloupka09faster, doyen09faster} to $O(|V| \cdot |E| \cdot W)$,
which is slightly better than the complexity of KASI.

The shortcoming of VI is that it takes enormous time before the vertices
with infinite $\lb^\Gamma$ and $\lwub_b^\Gamma$ value are identified. That's
why we first compute the vertices with $\nu < 0$ by some fast MPG solving
algorithm and then apply VI on the rest of the graph. For the lower-bound
problem, the vertices with $\nu < 0$ are exactly the vertices with infinite
$\lb^\Gamma$ value. For the lower-weak-upper-bound problem, the vertices
with $\nu < 0$ might be a strict subset of the vertices with infinite
$\lwub_b^\Gamma$ value, but still the preprocessing sometimes saves a lot of
time in practice. It is obvious that on MPGs with all vertices with $\nu
\geq 0$ the preprocessing does not help at all. It is also not helpful for
the lower-weak-upper-bound problem for small bound $b$.

According to our experiments, partly published
in~\cite{chaloupka09parallel}, the fastest algorithms in practice for
dividing the vertices of an MPG into those with $\nu \geq 0$ and $\nu < 0$
are the algorithm of Bj\"{o}rklund and
Vorobyov~\cite{bjorklund07combinatorial} (BV) and the algorithm of
Schewe~\cite{schewe08optimal} (SW). The fact that they are the fastest does
not directly follow from~\cite{chaloupka09parallel}, because that paper
focuses on parallel algorithms and computation of the exact $\nu$ values.

The original algorithm BV is a sub-exponential randomized algorithm.  To
prove that the algorithm is sub-exponential, some restrictions had to be
imposed. If these restrictions are not obeyed, BV runs faster.  Therefore,
we decided not to obey the restrictions and use only the ``deterministic
part'' of the algorithm. We used only the modified BV algorithm in our
experimental study. We even improved the complexity of the deterministic
algorithm from $O(|V|^2 \cdot |E| \cdot W)$ to $O(|V| \cdot (|V| \cdot \log
|V| + |E|) \cdot W)$ using the same technique as for the improvement of the
complexity of KASI which is described in the full version of this
paper~\cite{brim10usingTR}. Since the results of the improved BV were
significantly better on all input instances included in our experimental
study, all results of BV in this paper are the results of the improved BV.

The complexity of SW is $O(|V|^2 \cdot (|V| \cdot \log |V| + |E|) \cdot
W)$. It might seem that this is in contradiction with the title of Schewe's
paper~\cite{schewe08optimal}, because if some algorithm is optimal, one would expect
that there are no algorithms with better complexity. However, the term
``optimal'' in the title of the paper refers to the strategy improvement
technique. SW is also a strategy improvement algorithm, and the strategy
improvement steps in SW are optimal in a certain sense.

We note that any algorithm that divides the vertices of an MPG into those
with $\nu \geq 0$ and those $\nu < 0$ can be used to solve the lower-bound
and the lower-weak-upper-bound problem with the help of binary search, but
it requires introduction of auxiliary edges and vertices into the input MPG
and repeated application of the algorithm.  According to our experiments, BV
and SW run no faster than KASI.  Therefore, solving the two problems by
repeated application of BV and SW would lead to higher runtimes than the
runtimes of KASI. If we use the reduction technique
from~\cite{bouyer08infinite}, then BV/SW has to be executed $\Theta(|V|
\cdot \log(|V|\cdot W))$ times to solve the lower-bound problem, and
$\Theta(|V| \cdot \log b)$ times to solve the lower-weak-upper-bound. That's
why we compared KASI only with the algorithm VI and the two combined
algorithms: VI + BV and VI + SW. The complexities of BV and SW exceed the
complexity of VI, and so the complexities of the combined algorithms are the
complexities of BV and SW.  

\subsection{Input MPGs}

We experimented with completely random MPGs as well as more structured
synthetic MPGs and MPGs modeling simple reactive systems. The
synthetic MPGs were generated by two generators, namely
\textsf{SPRAND}~\cite{cherkassky96shortest} and
\textsf{TOR}~\cite{cherkassky99negative}, downloadable
from~\cite{sprandtor}. The outputs of these generators are only
directed weighted graphs, and so we had to divide vertices between Max
and Min ourselves. We divided them uniformly at random.  The MPGs
modeling simple reactive systems we created ourselves.

{\sf SPRAND} was used to generate the ``rand$x$'' MPG family. Each of
these MPGs contains $|E| = x\cdot|V|$ edges and consist of a random
Hamiltonian cycle and $|E| - |V|$ additional random edges, with
weights chosen uniformly at random from $[1, 10000]$. To make these
inputs harder for the algorithms, in each of them, we subtracted a
constant from each edge-weight so that the $\nu$ value of each vertex
is close to $0$.

{\sf TOR} was used for generation of the families ``sqnc'', ``lnc'', and
``pnc''. The sqnc and lnc families are 2-dimensional grids with wrap-around,
while the pnc family contains layered networks embedded on a torus. We also
created subfamilies of each of the three families by adding cycles to the
graphs.  For more information on these inputs we refer you
to~\cite{georgiadis09experimental} or~\cite{chaloupka09parallel}. Like for
the {\sf SPRAND} generated inputs, we adjusted each {\sf TOR} generated MPG
so that the $\nu$ value of each vertex is close to $0$.

As for the MPGs modeling simple reactive systems, we created three
parameterized models. The first is called ``collect'' and models a
robot on a ground with obstacles which has to collect items occurring
at different locations according to certain rules. Moving and even
idling consumes energy, and so the robot has to return to its docking
station from time to time to recharge. By solving the lower-bound, or
the lower-weak-upper-bound problem for the corresponding MPG,
depending on whether there is some upper bound on the robot's energy,
we find out from which initial configurations the robot has a strategy
which ensures that it will never consume all of its energy outside the
docking station, and we also get some strategy which ensures it. For
each ``good'' initial configuration, we also find out the minimal
sufficient amount of initial energy. We note that the energy is not a
part of the states of the model. If it was, the problem would be much
simpler.  We could simply compute the set of states from which Min has
a strategy to get the play to a state where the robot has zero energy
and it is not in the docking station. However, making the energy part
of the states would cause an enormous increase in the number of states
and make the model unmanageable.

The second model is called ``supply'' and models a truck which
delivers material to various locations the selection of which is
beyond its control. The goal is to never run out of the material so
that the truck is always able to satisfy each delivery request. We
also want to know the minimal sufficient initial amount of the
material.

The third model is called ``taxi'' and models a taxi which transports
people at their request. Its operation costs money and the taxi also
earns money. The goal is to never run out of money, and we also want to
know the minimal sufficient initial amount of money.

To get MPGs of manageable size, the models are, of course, very simplified,
but they are still much closer to real world problems than the synthetic
MPGs.
\vspace{-0.5em}

\subsection{Results}
The experiments were carried out on a machine equipped with two
dual-core Intel$^{\tiny\textcircled{r}}$ Xeon$^{\tiny\textcircled{r}}$
2.00GHz processors and 16GB of RAM, running GNU/Linux kernel version
2.6.26.  All algorithms were implemented in C++ and compiled with GCC
4.3.2 with the ``-O2'' option.

\begin{table}[t!]
  \begin{center}
      \begin{tabular}{|ll|r|r|r|r||r|r|r|r|}
        \hline 
        \multicolumn{2}{|c|}{} & 
        \multicolumn{4}{|c||}{\scriptsize lower-bound} & 
        \multicolumn{4}{|c|}{\scriptsize lower-weak-upper-bound}\\
        \hline \multicolumn{2}{|c|}{\raisebox{-0.5ex}{\footnotesize MPG}} & 
        \raisebox{-0.5ex}{\footnotesize \hspace{0.2cm}VI\hspace{0.2cm}} & 
        \raisebox{-0.5ex}{\footnotesize VI + BV} & 
        \raisebox{-0.5ex}{\footnotesize VI + SW} & 
        \raisebox{-0.5ex}{\footnotesize KASI} & 
        \raisebox{-0.5ex}{\footnotesize \hspace{0.2cm}VI\hspace{0.2cm} } & 
        \raisebox{-0.5ex}{\footnotesize VI + BV} & 
        \raisebox{-0.5ex}{\footnotesize VI + SW} & 
        \raisebox{-0.5ex}{\footnotesize KASI}\\[1ex] \hline
        \hline \scriptsize sqnc01 & \scriptsize (262k 524k) &
        \scriptsize n/a &
        \scriptsize 31.22 &
        \scriptsize 55.40 &
        \scriptsize 17.83 &
        \scriptsize 13.28 &
        \scriptsize 19.41 &
        \scriptsize 43.54 &
        \scriptsize 9.06 \\
        \hline \scriptsize sqnc02 & \scriptsize (262k 524k) &
        \scriptsize n/a &
        \scriptsize 13.30 &
        \scriptsize 20.14 &
        \scriptsize 11.01 &
        \scriptsize 2.88 &
        \scriptsize 10.70 &
        \scriptsize 17.52 &
        \scriptsize 3.57 \\
        \hline \scriptsize sqnc03 & \scriptsize (262k 525k) &
        \scriptsize n/a &
        \scriptsize 3.18 &
        \scriptsize 3.54 &
        \scriptsize 1.58 &
        \scriptsize 0.75 &
        \scriptsize 3.20 &
        \scriptsize 3.55 &
        \scriptsize 1.07 \\
        \hline \scriptsize sqnc04 & \scriptsize (262k 532k) &
        \scriptsize n/a &
        \scriptsize 9.34 &
        \scriptsize 11.49 &
        \scriptsize 8.48 &
        \scriptsize 1.65 &
        \scriptsize 8.55 &
        \scriptsize 10.70 &
        \scriptsize 2.53 \\
        \hline \scriptsize sqnc05 & \scriptsize (262k 786k) &
        \scriptsize n/a &
        \scriptsize 10.45 &
        \scriptsize 14.24 &
        \scriptsize 4.89 &
        \scriptsize 1.20 &
        \scriptsize 10.17 &
        \scriptsize 13.95 &
        \scriptsize 1.72 \\
        \hline \scriptsize lnc01 & \scriptsize (262k 524k) &
        \scriptsize 60.79 &
        \scriptsize 67.89 &
        \scriptsize 111.32 &
        \scriptsize 11.31 &
        \scriptsize 17.49 &
        \scriptsize 27.85 &
        \scriptsize 71.19 &
        \scriptsize 5.99 \\
        \hline \scriptsize lnc02 & \scriptsize (262k 524k) &
        \scriptsize 57.63 &
        \scriptsize 63.99 &
        \scriptsize 93.89 &
        \scriptsize 10.34 &
        \scriptsize 14.68 &
        \scriptsize 24.04 &
        \scriptsize 53.87 &
        \scriptsize 5.06 \\
        \hline \scriptsize lnc03 & \scriptsize (262k 525k) &
        \scriptsize n/a &
        \scriptsize 3.30 &
        \scriptsize 4.39 &
        \scriptsize 1.48 &
        \scriptsize 0.73 &
        \scriptsize 3.34 &
        \scriptsize 4.41 &
        \scriptsize 1.03 \\
        \hline \scriptsize lnc04 & \scriptsize (262k 528k) &
        \scriptsize n/a &
        \scriptsize 21.05 &
        \scriptsize 25.28 &
        \scriptsize 10.63 &
        \scriptsize 3.53 &
        \scriptsize 11.65 &
        \scriptsize 15.64 &
        \scriptsize 4.24 \\
        \hline \scriptsize lnc05 & \scriptsize (262k 786k) &
        \scriptsize n/a &
        \scriptsize 10.89 &
        \scriptsize 11.30 &
        \scriptsize 4.77 &
        \scriptsize 1.17 &
        \scriptsize 10.64 &
        \scriptsize 11.03 &
        \scriptsize 1.65 \\
        \hline \scriptsize pnc01 & \scriptsize (262k 2097k) &
        \scriptsize n/a &
        \scriptsize 24.27 &
        \scriptsize 16.08 &
        \scriptsize 3.80 &
        \scriptsize 1.41 &
        \scriptsize 24.31 &
        \scriptsize 16.08 &
        \scriptsize 1.98 \\
        \hline \scriptsize pnc02 & \scriptsize (262k 2097k) &
        \scriptsize n/a &
        \scriptsize 25.49 &
        \scriptsize 15.37 &
        \scriptsize 3.80 &
        \scriptsize 1.43 &
        \scriptsize 25.55 &
        \scriptsize 15.38 &
        \scriptsize 1.98 \\
        \hline \scriptsize pnc03 & \scriptsize (262k 2098k) &
        \scriptsize n/a &
        \scriptsize 23.48 &
        \scriptsize 17.66 &
        \scriptsize 3.86 &
        \scriptsize 1.48 &
        \scriptsize 23.53 &
        \scriptsize 17.66 &
        \scriptsize 2.04 \\
        \hline \scriptsize pnc04 & \scriptsize (262k 2101k) &
        \scriptsize n/a &
        \scriptsize 26.36 &
        \scriptsize 25.24 &
        \scriptsize 3.91 &
        \scriptsize 1.49 &
        \scriptsize 26.34 &
        \scriptsize 25.23 &
        \scriptsize 2.05 \\
        \hline \scriptsize pnc05 & \scriptsize (262k 2359k) &
        \scriptsize n/a &
        \scriptsize 27.09 &
        \scriptsize 29.69 &
        \scriptsize 4.71 &
        \scriptsize 1.97 &
        \scriptsize 27.15 &
        \scriptsize 29.69 &
        \scriptsize 2.51 \\
        \hline \scriptsize rand5 & \scriptsize (262k 1310k) &
        \scriptsize n/a &
        \scriptsize 19.16 &
        \scriptsize 20.42 &
        \scriptsize 4.55 &
        \scriptsize 1.65 &
        \scriptsize 19.27 &
        \scriptsize 20.54 &
        \scriptsize 2.39 \\
        \hline \scriptsize rand5b & \scriptsize (524k 2621k) &
        \scriptsize n/a &
        \scriptsize 36.29 &
        \scriptsize 33.06 &
        \scriptsize 10.09 &
        \scriptsize 3.53 &
        \scriptsize 36.52 &
        \scriptsize 33.24 &
        \scriptsize 5.17 \\
        \hline \scriptsize rand5h & \scriptsize (1048k 5242k) &
        \scriptsize n/a &
        \scriptsize 86.55 &
        \scriptsize 59.01 &
        \scriptsize 21.35 &
        \scriptsize 7.45 &
        \scriptsize 87.16 &
        \scriptsize 59.48 &
        \scriptsize 11.07 \\
        \hline \scriptsize rand10 & \scriptsize (262k 2621k) &
        \scriptsize n/a &
        \scriptsize 39.30 &
        \scriptsize 36.96 &
        \scriptsize 5.60 &
        \scriptsize 2.37 &
        \scriptsize 39.39 &
        \scriptsize 37.00 &
        \scriptsize 3.68 \\
        \hline \scriptsize rand10b & \scriptsize (524k 5242k) &
        \scriptsize n/a &
        \scriptsize 105.69 &
        \scriptsize 43.43 &
        \scriptsize 14.54 &
        \scriptsize 5.07 &
        \scriptsize 105.97 &
        \scriptsize 43.45 &
        \scriptsize 7.98 \\
        \hline \scriptsize rand10h & \scriptsize (1048k 10485k) &
        \scriptsize n/a &
        \scriptsize 140.46 &
        \scriptsize 110.68 &
        \scriptsize 29.27 &
        \scriptsize 11.38 &
        \scriptsize 140.57 &
        \scriptsize 110.82 &
        \scriptsize 17.52 \\
        \hline \scriptsize collect1 & \scriptsize (636k 3309k) &
        \scriptsize 996.08 &
        \scriptsize 1027.12 &
        \scriptsize 1032.55 &
        \scriptsize 5.68 &
        \scriptsize 531.40 &
        \scriptsize 544.77 &
        \scriptsize 563.78 &
        \scriptsize 4.89 \\
        \hline \scriptsize collect2 & \scriptsize (636k 3309k) &
        \scriptsize 338.56 &
        \scriptsize 352.45 &
        \scriptsize 367.12 &
        \scriptsize 5.70 &
        \scriptsize 181.35 &
        \scriptsize 189.17 &
        \scriptsize 208.52 &
        \scriptsize 4.89 \\
        \hline \scriptsize supply1 & \scriptsize (363k 1014k) &
        \scriptsize 6956.23 &
        \scriptsize 16.03 &
        \scriptsize 109.72 &
        \scriptsize 1.79 &
        \scriptsize 7.72 &
        \scriptsize 8.87 &
        \scriptsize 102.97 &
        \scriptsize 1.85 \\
        \hline \scriptsize supply2 & \scriptsize (727k 2030k) &
        \scriptsize 28046.54 &
        \scriptsize 65.08 &
        \scriptsize 449.47 &
        \scriptsize 3.64 &
        \scriptsize 30.84 &
        \scriptsize 33.31 &
        \scriptsize 418.88 &
        \scriptsize 3.77 \\
        \hline \scriptsize taxi1 & \scriptsize (509k 979k) &
        \scriptsize 11.64 &
        \scriptsize 12.85 &
        \scriptsize 13.16 &
        \scriptsize 1.29 &
        \scriptsize 0.70 &
        \scriptsize 1.49 &
        \scriptsize 2.17 &
        \scriptsize 1.38 \\
        \hline \scriptsize taxi2 & \scriptsize (509k 979k) &
        \scriptsize 6.00 &
        \scriptsize 7.03 &
        \scriptsize 7.51 &
        \scriptsize 1.29 &
        \scriptsize 0.70 &
        \scriptsize 1.49 &
        \scriptsize 2.17 &
        \scriptsize 1.38 \\
        \hline
    \end{tabular}
  \end{center}
\vspace{-0.5em}
  \caption{Runtimes of the experiments (in seconds)}\label{tab:xp}
\vspace{-0.5em}
\end{table}

Table~\ref{tab:xp} gives the results of our experiments. The first column of
the table contains names of the input MPGs. Numbers of vertices and edges,
in thousands, are in brackets. The MPGs prefixed by ``sqnc'', ``lnc'', and
``pnc'' were generated by the {\sf TOR} generator. They all contain $2^{18}$
vertices. The MPGs prefixed by ``rand'' were generated by the {\sf SPRAND}
generator. Both for the rand5 and rand10 family, we experimented with three
sizes of graphs, namely, with $2^{18}$ vertices -- no suffix, with $2^{19}$
vertices -- suffix ``b'', and with $2^{20}$ vertices -- suffix
``h''. Finally, the MPGs prefixed by ``collect'', ``supply'', and ``taxi''
are the models of simple reactive systems created by ourselves. For each
model, we tried two different values of parameters.

Each MPG used in the experiments has eight columns in the table. Each
column headed by a name of an algorithm contains execution times of
that algorithm in seconds, excluding the time for reading input. The
term ``n/a'' means more than 10 hours. The first four columns contain
results for the lower-bound problem, the last four columns contain the
results for the lower-weak-upper-bound problem, which contains a bound
$b$ as a part of the input. If the bound is too high, the algorithms
essentially solve the lower-bound problem, and so the runtimes are
practically the same as for the lower-bound problem. If the bound is
too low, all vertices in our inputs have infinite $\lwub_b^\Gamma$
value, and they become very easy to solve. We tried various values of
$b$, and for this paper, we selected as $b$ the average $\lb^\Gamma$
value of the vertices with finite $\lb^\Gamma$ value divided by $2$,
which seems to be a reasonable amount so that the results provide
insight. We note that smaller $b$ makes the computation of VI and KASI
faster.  However, the BV and SW parts of VI + BV and VI + SW always
perform the same work, and so for $b \ll (|V| - 1)\cdot W$, the
combined algorithms are often slower than VI alone.

The table shows that the algorithm KASI was the fastest on all inputs for
the lower-bound problem. For the lower-weak-upper-bound problem it was never
slower than the fastest algorithm by more than a factor of $2$, and for some
inputs it was significantly faster. This was true for all values of $b$ that
we tried. Therefore, the results clearly suggest that KASI is the best
algorithm. In addition, there are several other interesting points.

VI is practically unusable for solving the lower-bound problem for MPGs with
some vertices with $\nu < 0$. Except for lnc01--02, collect1--2, and
taxi1--2, all input MPGs had vertices with $\nu < 0$. The preprocessing by
BV and SW reduces the execution time by orders of magnitude for these
MPGs. On the other hand, for the lower-weak-upper-bound problem for the
bound we selected, VI is often very fast and the preprocessing slows the
computation down in most cases. VI was even faster than KASI on a lot of
inputs. However, the difference was never significant, and it was mostly
caused by the initialization phase of the algorithms, which takes more time
for the more complex algorithm KASI. Moreover, for some inputs, especially
from the ``collect'' family, VI is very slow. VI makes a lot of iterations
for the inputs from the collect family, because the robot can survive for
quite long by idling, which consumes a very small amount of energy per time
unit. However, it cannot survive by idling forever. The $i$-th iteration of
VI computes the minimal sufficient initial energy to keep the energy level
non-negative for $i$~time units, and so until the idling consumes at
least as much energy as the minimal sufficient initial energy to keep the
energy level non-negative forever, new iterations have to be started. We
believe that this is a typical situation for this kind of application.
Other inputs for which VI took a lot of time are: sqnc01, lnc01--02,
supply1--2.

Finally, we comment on scalability of the algorithms. As the experiments on
the {\sf SPRAND} generated inputs suggest, the runtimes of the algorithms
increase no faster than the term $|V|\cdot|E|$, and so they are able to
scale up to very large MPGs.

\vspace{-0.5em}
\section{Conclusion}
\vspace{-0.38em}

We proposed a novel algorithm for solving the lower-bound and the
lower-weak-upper-bound problems for MPGs. Our algorithm, called Keep Alive
Strategy Improvement (KASI), is based on the strategy improvement technique
which is very efficient in practice. To demonstrate that the algorithm is
able to solve the two problems for large MPGs, we carried out an
experimental study. In the study we compared KASI with the value iteration
algorithm (VI) from~\cite{chaloupka09faster, doyen09faster}, which we also improved by
combining it with the algorithm of Bj\"{o}rklund and
Vorobyov~\cite{bjorklund07combinatorial} (BV) and the algorithm of Schewe
(SW). KASI is the clear winner of the experimental study.

Two additional results of this paper are the improvement of the complexity
of BV, and characterization of Min's optimal strategies w.r.t. the
lower-weak-upper-bound problem.
\vspace{-0.66em}

\bibliography{citations}

\end{document}